\documentclass{article}
\usepackage{palatino,amsfonts}
\usepackage[dvips]{graphicx}
\usepackage{amsmath}
\usepackage{amssymb}

\title{Finding the ``truncated'' polynomial that is closest to a
function}

\author{Nicolas Brisebarre \footnote{LArAl, Universit\'e Jean Monnet,
           23, rue du Dr P. Michelon, F-42023 Saint-\'Etienne Cedex,
            France and  LIP/Ar\'enaire
  (CNRS-ENS Lyon-INRIA-UCBL),
46 All\'ee d'Italie, F-69364 Lyon Cedex 07 {\sc France},
Nicolas.Brisebarre@ens-lyon.fr}\,   and Jean-Michel Muller
\footnote{LIP/Ar\'enaire (CNRS-ENS Lyon-INRIA-UCBL), 46 All\'ee
d'Italie, F-69364 Lyon Cedex 07 {\sc France},
Jean-Michel.Muller@ens-lyon.fr}}

\newtheorem{property}{Property}
\newtheorem{definition}{Definition}
\newtheorem{prop}{Proposition}
\newtheorem{cor}{Corollary}
\newtheorem{theorem}{Theorem}
\newtheorem{rem}{Remark}
\newcommand{\ree}{\mathbb R}

\begin{document}
\maketitle

\begin{abstract}
When implementing regular enough functions (e.g., elementary or
special functions) on a computing system, we frequently use
polynomial approximations. In most cases, the polynomial that best
approximates (for a given distance and in a given interval) a
function has coefficients that are not exactly representable with
a finite number of bits. And yet, the polynomial approximations
that are actually implemented do have coefficients that are
represented with a finite - and sometimes small - number of bits:
this is due to the finiteness of the floating-point
representations (for software implementations), and to the need to
have small, hence fast and/or inexpensive, multipliers (for
hardware implementations). We then have to consider polynomial
approximations for which the degree-$i$ coefficient has at most
$m_i$ fractional bits (in other words, it is a rational number
with denominator $2^{m_i}$). We provide a general method for
finding the best polynomial approximation under this constraint.
Then, we suggest refinements than can be used to accelerate our
method.
\end{abstract}

\section*{Introduction}
{\sl All the functions considered in this article are real valued
functions of the real variable and all the polynomials have real
coefficients.} After an initial \emph{range reduction}
step~\cite{PH83,Ng92,DMMM95}, the problem of evaluating a function
$\varphi$ in a large domain on a computer system is reduced to the
problem of evaluating a possibly different function $f$ in a small
domain, that is generally of the form $[0,a]$, where $a$ is a
small nonnegative real. Polynomial approximations are among the
most frequently chosen ways of performing this last approximation.

Two kinds of polynomial approximations are used: the
approximations that minimize the ``average error,'' called {\em
least squares approximations\/}, and the approximations that
minimize the worst-case error, called {\em least maximum
approximations\/}, or {\em minimax approximations\/}.  In both
cases, we want to minimize a distance $||p-f||$, where $p$ is a
polynomial of a given degree. For least squares approximations,
that distance is:

$$
||p-f||_{2,[0,a]} = \left(\int_{0}^{a}w(x)\left(f(x)-p(x)\right)^2
dx\right)^{1/2},
$$

\noindent where $w$ is a continuous {\em weight function\/}, that
can be used to  select parts of $[0,a]$ where we want the
approximation to be more accurate. For minimax approximations, the
distance is:

$$
|| p -f ||_{\infty,[0,a]} = \max_{0 \leq x \leq a} |p(x)-f(x)|.
$$

The least squares approximations are computed by a projection
method using orthogonal polynomials. Minimax approximations are
computed using an algorithm due to Remez~\cite{Rem34,Har68}.
See~\cite{Mul97,Mar2000} for recent presentations of elementary
function algorithms.

In this paper, we are concerned with minimax approximations. Our
approximations will be used in finite-precision arithmetic. Hence,
the computed polynomial coefficients are usually rounded: the
coefficient $p_i$ of the minimax approximation

$$
p(x) = p_0 + p_1x + \cdots{} + p_nx^n
$$

\noindent is rounded to, say, the nearest multiple of $2^{-m_i}$.
By doing that, we obtain a slightly different polynomial
approximation $\hat{p}$. But \emph{we have no guarantee that
$\hat{p}$ is the best minimax approximation to $f$ among the
polynomials whose degree $i$ coefficient is a multiple of
$2^{-m_i}$.} The aim of this paper is to give a way of finding
this ``best truncated approximation''. We have two goals in mind:

\begin{itemize}
   \item rather low precision (say, around $15$ bits),
   hardware-oriented, for specific-purpose implementations.
   In such cases, to minimize multiplier sizes (which increases speed
   and saves silicon area), the values of
   $m_i$, for $i \geq 1$, should be very small. The degrees of the
   polynomial approximations are low. Typical recent examples are given
   in~\cite{CaoWeiChe2001,PinBrugMul2001}. Roughly speaking, what
   matters here is to
   reduce the cost (in terms of delay and area) without making the
   accuracy unacceptable;

   \item single-precision or double-precision, software-oriented,
   general-purpose implementations for implementation on current
  microprocessors.
   Using Table-driven methods, such as the ones suggested by
   Tang~\cite{Tan89,Tan90,Tan91,Tan92}, the degree of the
   polynomial approximations can be made rather low. Roughly speaking,
  what matters in that case is to get very high accuracy, without
   making the cost (in terms of delay and memory) unacceptable.
\end{itemize}

The outline of the paper is the following. We give an account of
Chebyshev polynomials and some of their properties in Section
\ref{seccheb}. Then, in Section \ref{secget}, we provide a general
method that finds the ``best truncated approximation'' of a
function $f$ over a compact  interval $[0,a]$. We finish with two
examples.

Our method is implemented in Maple programs that can be downloaded
from {\tt http://www.ens-lyon.fr/\~{ }nbriseba/trunc.html}. We
plan to prepare  a C version of these programs which should be
much faster.

\section{Some reminders on Chebyshev polynomials} \label{seccheb}

\begin{definition}[Chebyshev polynomials]
The Chebyshev polynomials can be defined either by the recurrence
relation

\begin{equation}
\label{def1-Chebyshev}
 \left\{
        \begin{array}{lll}
                T_0(x) & = & 1  \\
                T_1(x) & = & x \\
                T_n(x) & = & 2xT_{n-1}(x)-T_{n-2}(x);
        \end{array}
        \right.
\end{equation}

or by

\begin{equation}
\label{def2-Chebyshev}
 T_n(x) = \left\{
 \begin{array}{ll} \cos\left(n
\cos^{-1} x\right) & (|x| \leq 1) \\
\cosh\left(n \cosh^{-1} x\right) & (x > 1).
\end{array}\right.
\end{equation}
\end{definition}

The first Chebyshev polynomials are listed below.

$$
\begin{array}{lll}
T_0(x) & = & 1,  \\
T_1(x) & = & x, \\
T_2(x) &=& 2x^2-1, \\
T_3(x) &=& 4x^3-3x, \\
T_4(x) &=& 8x^4-8x^2+1, \\
T_5(x) &=& 16x^5-20x^3+5x.
\end{array}
$$

\begin{figure}
\includegraphics[scale=0.5,angle=-90]{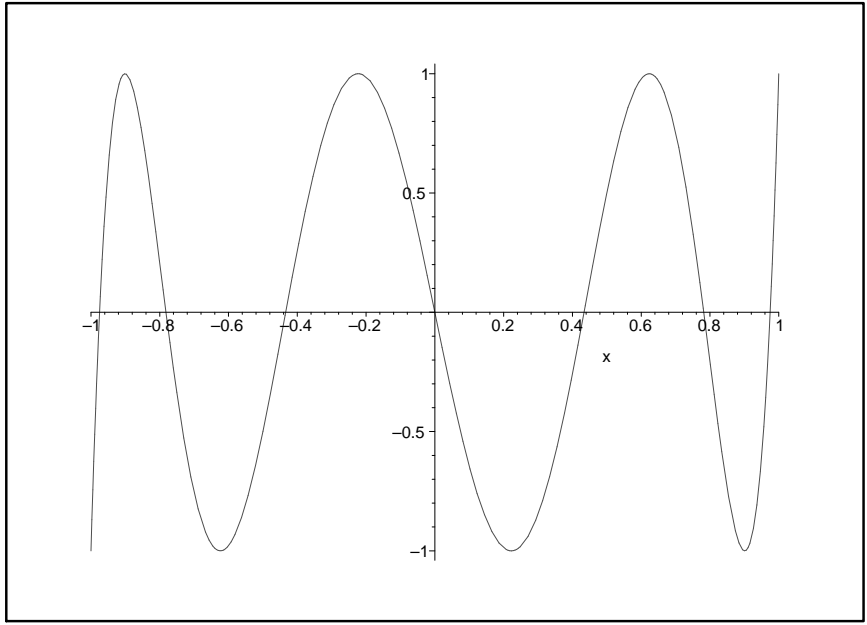}
 \caption{Graph of the polynomial $T_7(x).$}\label{T7plot}
\end{figure}

An example of Chebyshev polynomial ($T_7$) is plotted in
Fig.~\ref{T7plot}.

These polynomials play a central role in approximation theory.
Among their many properties, the following ones will be useful in
the sequel of this paper. A presentation of the Chebyshev
polynomials can be found in~\cite{bor} and especially in
\cite{Riv1990}.

\begin{property} For $n \geq 0$, we have
$$
T_n (x) = \frac{n}{2} \sum_{k = 0}^{\lfloor n/2 \rfloor} (-1)^k
\frac{(n-k-1)!}{k! (n-2k)!}  (2x)^{n -2k}.
$$

Hence,  $T_n$ has degree $n$ and  %Its degree-$n$
its leading  coefficient is $2^{n-1}$. It has $n$ real roots, all
strictly between $-1$ and $1$.
\end{property}

\begin{property}
\label{extremaTn} There are exactly $n+1$ values $x_0, x_1, x_2,
\ldots, x_n$ such that
$$-1 = x_0 < x_1 < x_2 < \cdots{} < x_n = 1,$$
which satisfy
$$
T_n(x_i) = (-1)^{n-i} \max_{x \in [-1,1]} |T_n(x)| \quad \forall
i, \, i =0, \ldots, \, n.
$$
%and
%$$
%T_n(x_{i+1}) = -T_n(x_i).
%$$
That is, the maximum absolute value of $T_n$ is attained at the
$x_i$'s, and the sign of $T_n$ alternates at these points.
\end{property}

We recall  that a \emph{monic}
polynomial is a polynomial whose leading coefficient %of highest

                                                     %degree
is $1$.

\begin{property}[Monic polynomials of smallest norm]
\label{monic} Let $a, \, b \in \ree$, $ a \leq b$. The monic
degree-$n$ polynomial having the smallest $||.||_{\infty,[a,b]}$
norm in $[a,b]$ is
$$
\frac{(b-a)^n}{2^{2n-1}} T_n\left(\frac{2x-b-a}{b-a}\right).
$$
\end{property}

The central result in polynomial approximation theory is the
following theorem, due to Chebyshev.

\begin{theorem}[Chebyshev]
\protect\label{chebyshev_Theorem}\index{Chebyshev!theorem}  Let
$a, \, b \in \ree$, $ a \leq b$. The polynomial $p$ is the minimax
approximation of degree $\leq n$ to a continuous function $f$ on
$[a,b]$ if and only if there exist at least $n+2$ values
$$
a \leq x_0 < x_1 < x_2 < \ldots{} < x_{n+1} \leq b
$$
such that:
$$
p(x_i)-f(x_i) = (-1)^i\left[p(x_0)-f(x_0)\right] = \pm
||f-p||_{{\infty,[a,b]}}.
$$
\end{theorem}

Throughout the paper, we will make frequent use of the polynomials
$$
T_n^*(x) = T_n(2x-1).
$$

The first polynomials $T_n^*$ are given below. We have (see
\cite[Chap. 3]{fp} for example) $T_n^* (x) = T_{2n} (x^{1/2})$,
hence all the coefficients of $T_n^*$ are nonzero integers.

$$
\begin{array}{lll}
T_0^*(x) &=& 1, \\
T_1^*(x) &=& 2x-1, \\
T_2^*(x) &=& 8x^2 - 8x + 1, \\
T_3^*(x) &=& 32x^3 - 48x^2 + 18x -1, \\
T_4^*(x) &=& 128x^4 - 256x^3 + 160x^2 - 32x + 1,\\
T_5^*(x) &=& 512x^5 - 1280x^4 + 1120 x^3 - 400 x^2 + 50x - 1.\\
\end{array}
$$

\begin{theorem}[Polynomial of smallest norm with degree-$k$
coefficient equal to $1$.] \label{smallest-i} Let $ a \in
(0,+\infty )$, define
$$\beta_0 + \beta_1x + \beta_2x^2 +\cdots{} + \beta_nx^n =
T_n^*\left(\frac{x}{a}\right).$$ Let $k$ be an integer, $0 \leq k
\leq n$, the polynomial
$$
\frac{1}{\beta_k} T_n^*\left(\frac{x}{a}\right).
$$
has the smallest $||.||_{\infty,[0,a]}$ norm in $[0,a]$ among the
polynomials of degree at most $n$ with a degree-$k$ coefficient
equal to $1$. That norm is $|1/\beta_k|$.

Moreover, when $k = n = 0$ or $1 \leq k \leq n$, this polynomial
is the only one having this property.

\end{theorem}

Proving this theorem first requires the following results.

\begin{prop} \label{muntz}
Let  $(\delta_i)_{i =0,\ldots,n}$ be an increasing sequence of
nonnegative integers and
$$
P(x)  = a_0 x^{\delta_0} + \cdots + a_n x^{\delta_n} \in \ree [x],
$$
then either $P =0$ or $P$ has at most $n$ zeros  in $(0,+\infty)$.
\end{prop}
\noindent \textbf{Proof.} % of Proposition~\ref{muntz}}
By induction on $n$. For $n = 0$, it is straightforward. Now we
  assume that the property is true until the rank $n$. Let $P(x)  =
  a_0 x^{\delta_0} + \cdots + a_n x^{\delta_n} +  a_{n+1} x^{\delta_{n+1}} \in
  \ree [x]$  with $0 \leq \delta_0 < \cdots < \delta_{n+1}$ and $a_0
  a_1 \ldots a_{n+1} \neq 0$. Assume that $P$
  has at least $n+2$ zeros in $(0,+\infty)$. Then  $P_1 =
  P/x^{\delta_0}$ has at least   $n+2$ zeros in $(0,+\infty)$.

Thus, the nonzero polynomial  $P_1' (x) = (\delta_1
  -\delta_0) a_1 x^{\delta_1 -\delta_0} + \cdots +  (\delta_{n+1}
  -\delta_0) a_{n+1} x^{\delta_{n+1} -\delta_0} $
has, from Rolle's Theorem, at least $n+1$  zeros in $(0,+\infty)$,
which contradicts the induction hypothesis. $\Box$

\begin{cor} \label{cormuntz} Let $k$ be an integer, $1 \leq k
\leq n$,  and
$$
P(x) = \displaystyle\sum_{j = 0
 \atop j \neq k }^n e_j x^j  \in \ree [x].
$$
If $P$ has at least $n$ zeros in $[0,+\infty)$ and at most a
simple zero in $0$, then  $P =0$.
\end{cor}
\noindent \textbf{Proof.} If $P(0) \neq 0$, then $P$ has at least
 $n$ zeros in $(0,+\infty)$, hence $P =0$ from Proposition
 \ref{muntz}. Suppose now that $P(0) = 0$. We can rewrite
$P$ as  $P (x) = \displaystyle\sum_{j = 1
 \atop j \neq k }^n e_j x^j$. As $P$  has  at least $n-1$ zeros  in
 $(0,+\infty)$, it must yet
  vanish identically from Proposition \ref{muntz}. $\Box$

\noindent \textbf{Proof of Theorem~\ref{smallest-i}}. We give the
proof in the case $a=1$ (the general case is a straightforward
generalization).

The case $k=n=0$ is straightforward.

 Denote $T_n^* (x) = \displaystyle
\sum_{k=0}^n a_k x^k$.  From Property \ref{extremaTn}, there exist
$0 = \eta_0 < \eta_1 < \cdots < \eta_n =1$ such that

$$
%a_k^{-1} T_n^* (x) = x^k +\displaystyle \sum_{j = 0 \atop
%j   \neq k }^n c_j x^j \quad \mbox{ and } \quad
a_k^{-1} T_n^* (\eta_i) = a_k^{-1} (-1)^{n-i}  \left \Vert T_n^*
\right \Vert_{\infty,[0,1]} =  a_k^{-1} (-1)^{n-i}.$$

Now, we assume $1 \leq k \leq n$. This part of the proof follows
step by step the proof of Theorem 2.1 in \cite{Riv1990}.
 Let $q(x) = \displaystyle\sum_{j = 0, \atop j \neq k}^n c_j x^j  \in \ree [x]$ satisfy  $\Vert x^k - q(x)
\Vert_{\infty,[0,1]} \leq |a_k^{-1}|$. We suppose that $x^k - q
\neq a_k^{-1} T_n^*$. Then the polynomial $P(x) = a_k^{-1} T_n^*
(x) - (x^k -q(x)) $ has the form $ \displaystyle\sum_{j = 0, \atop
j \neq k }^n d_j x^j $ and is not identically zero.

Hence there exist  $i$ and $j$, $0 \leq i < j \leq n$, such that
$P (\eta_0) = \cdots = P (\eta_{i-1}) =0$, $P(\eta_i) \neq 0$ and
$ P(\eta_j) \neq 0$,  $P (\eta_{j+1}) = \cdots = P (\eta_{n}) =0$
(otherwise, the nonzero
 polynomial $P$ would have at least $n$ distinct roots in $[0,1]$
 which would contradict Corollary \ref{cormuntz}). Let $l $
such that $ P(\eta_l) \neq 0$ then sgn $P(\eta_l) = $ sgn
$a_k^{-1} T_n^* (\eta_l) = (-1)^{n-l}$ sgn $a_k^{-1}$. Let $m$
such that $ P(\eta_l) \neq 0$, $ P(\eta_{l+1}) = \cdots = P
(\eta_{l+m -1}) = 0$, $ P(\eta_{l+m}) \neq 0$ : $P$ has at least
$m-1$ zeros in $[\eta_l,
  \eta_{l+m}]$. We distinguish two cases:

\begin{itemize}
\item If $m$ is even, we have sgn $P(\eta_l) =$ sgn
$P(\eta_{l+m})$  and
   thus, $P$ must have an even number of zeros (counted with
   multiplicity) in $[\eta_l,
  \eta_{l+m}]$.

\item If $m$ is odd, we have sgn $P(\eta_l) = -$ sgn
$P(\eta_{l+m})$  and
   thus, $P$ must have an odd number of zeros (counted with
   multiplicity) in $[\eta_l, \eta_{l+m}]$.
\end{itemize}

In both cases, we conclude that $P$ has at least $m$ zeros in
$[\eta_l,
 \eta_{l+m}]$.

Then $P$ has at least $j-i$ zeros in $[\eta_i,\eta_j]$. Finally,
$P$ has not less than $i + (j-i) + n- j = n$ zeros in $[0,1]$ ($P$
has at least $i$ zeros in $[\eta_0,\eta_i)$ and at least $n-j$
zeros in $(\eta_j,
  \eta_n]$). Note that we also obtained that $P$ has  no less than
$n-1$ zeros in $(0,1]$. As $P$ is nonzero, this contradicts
 Corollary~\ref{cormuntz}. % that $P$ vanishes identically.

To end, we assume $k =0$ and $n \geq 1$. Let $q(x) =
 \displaystyle\sum_{j = 1}^n c_j x^j  \in \ree [x]$
 satisfy  $\Vert 1 - q(x)
\Vert_{\infty,[0,1]} < |a_0^{-1}|$. Then the polynomial $P(x) =
 a_0^{-1} T_n^*(x) - (1 -q(x)) $ has the form $ \displaystyle\sum_{j = 1 }^n d_j x^j
$ and is not identically zero. This polynomial changes sign
between any two consecutive extrema of $T_n^* $, hence it has at
least  $n$ zeros in $(0,1)$. As it cancels also at $0$, we deduce
that $P$ vanishes identically, which is the contradiction desired.
$\Box$

\begin{rem} When $k =0$ and $n \geq 1$, it is not possible to prove
 unicity:
%there may be other polynomials
% having the smallest $||.||_{\infty,[0,a]}$ norm in $[0,a]$ among the
%polynomials of degree at most $n$ with a constant coefficient
%equal to $1$.
for example, let $a =1$, $k=0$, $n = 1$, the polynomials $1-
\lambda x$ with $\lambda \in [0,2]$ have all a
$||.||_{\infty,[0,1]}$ norm equal to $1$.
\end{rem}

\section{Getting the ``truncated'' polynomial that is
  closest to a function in $[0,a]$}\label{secget}
Let $a \in (0,+\infty)$, let $f$ be a function defined on $[0,a]$
and $m_0$,
 $m_1$, \ldots{}, $m_n$  be $n+1$ integers. Define ${\cal
P}_n^{[m_0,m_1,\ldots{},m_n]}$ as the set of the polynomials of
degree less than or equal to $n$ whose degree-$i$ coefficient is a
multiple of $2^{-m_i}$ for all $i$ between $0$ and $n$ (we will
call these polynomials ``truncated polynomials'').

Let $p$ be the minimax approximation to $f$ on $[0,a]$. Define
$\hat{p}$ as the polynomial whose degree-$i$ coefficient is
obtained by rounding the degree-$i$ coefficient of $p$ to the
nearest multiple of $2^{-m_i}$ (with an arbitrary choice in case
of a tie) for $i = 0, \ldots, \, n$:
 $\hat{p}$ is an element of ${\cal  P}_n^{[m_0,m_1,\ldots{},m_n]}$.

 Also define ${\epsilon}$ and $\hat{\epsilon}$ as

$$
\epsilon =   ||f-p||_{\infty,[0,a]}\,  \mbox{ and } \,
\hat{\epsilon} = ||f - \hat{p}||_{\infty,[0,a]}.
$$

We assume that $\hat{\epsilon} \neq 0$. Let $\lambda \in \left
[\frac{\epsilon}{\hat{\epsilon}} , 1 \right ]$, we are looking for
a truncated polynomial $p^{\star} \in {\cal
P}_n^{[m_0,m_1,\ldots{},m_n]}$ such that

\begin{equation*}
||f-p^{\star}||_{\infty,[0,a]} = \min_{q \in {\cal
P}_n^{[m_0,m_1,\ldots{},m_n]}} ||f-q||_{\infty,[0,a]}
\end{equation*}

and

\begin{equation}\label{inflambda}
||f-p^{\star}||_{\infty,[0,a]} \leq \lambda
||f-\hat{p}||_{\infty,[0,a]}.
\end{equation}

When $\lambda =1$, this problem has  a solution since $\hat{p}$
 satisfies \eqref{inflambda}.
 It should be noticed that, in that case, $p^{\star}$ is
not necessarily equal to $\hat{p}$.% (in most practical cases, it is not).

In the following, we compute bounds on the coefficients of a
polynomial $q \in {\cal P}_n^{[m_0,m_1,\ldots{},m_n]}$ such that
if $q$ is not within these bounds, then
$$
|| p - q ||_{\infty,[0,a]} > \epsilon + \lambda \hat{\epsilon}.
$$

Knowing these bounds will allow an exhaustive searching of
$p^{\star}$. To do that, consider a polynomial $q$ whose
degree-$i$ coefficient is $p_i + \delta_i$, with $\delta_i \neq
0$. Let us see how close can $q$ be to $p$. We  have

$$
(q-p)(x) = \delta_i x^i + \sum_{0 \leq j \leq n, \atop j \neq i}
(q_j-p_j)x^j.
$$

Hence, $||q-p||_{\infty,[0,a]}$ is minimum implies that
$$
||x^i + \frac{1}{\delta_i} \sum_{0 \leq j \leq n, \atop j \neq i}
(q_j-p_j)x^j||_{\infty,[0,a]}
$$
is minimum.

Hence, we have to find the polynomial of degree $n$, with fixed
degree-$i$ coefficient, whose norm is smallest. This is given by
Theorem~\ref{smallest-i}. Therefore, we have
$$
||x^i + \frac{1}{\delta_i} \sum_{0 \leq j \leq n, \atop  j \neq i}
(q_j-p_j)x^j||_{\infty,[0,a]} \geq \frac{1}{|\beta_i|},
$$
where $\beta_i$ is the nonzero degree-$i$ coefficient of
$T_n^*(x/a)$. Therefore, we must have

$$
||q-p||_{\infty,[0,a]} \geq \frac{\delta_i}{|\beta_i|}.
$$

Now,
%since $||f- \hat{p}||_{\infty,[0,a]} = \hat{\epsilon}$, and
%since $\hat{p} \in {\cal P}_n^{[m_0,m_1,\ldots{},m_n]}$,
if a polynomial is at a distance greater than $\epsilon + \lambda
\hat{\epsilon}$ from $p$, it cannot be $p^{\star}$ since
$$
||q - f||_{\infty,[0,a]} \geq || q -  p ||_{\infty,[0,a]} -
||p-f||_{\infty,[0,a]}  >  \lambda \hat{\epsilon}.
$$
 Therefore, if
there exists $i$, $0 \leq i \leq n$, such that
$$
|\delta_i| > (\epsilon + \lambda \hat{\epsilon})|\beta_i|
$$
then $ || q -  p || > \epsilon + \lambda \hat{\epsilon}$ and
therefore $ q \neq p^{\star}$.
%the polynomial $q$  cannot be the element of
% ${\cal P}_n^{[m_0,m_1,\ldots{},m_n]}$ that is closest to $f$.
 Hence, the
degree-$i$ coefficient of $p^{\star}$ necessarily lies in the
interval $[p_i- (\epsilon +\lambda\hat{\epsilon})|\beta_i|,p_i+
  (\epsilon +  \lambda\hat{\epsilon})|\beta_i|]$.
Thus we have

\begin{equation}\label{bounds}
\underbrace{\left\lceil 2^{m_i} (p_i - (\epsilon +
  \lambda\hat{\epsilon})|\beta_i|)
\right\rceil}_{m_i} \leq 2^{m_i} {p_i^{\star}} \leq
 \underbrace{\left\lfloor 2^{m_i}(p_i +
(\epsilon +  \lambda\hat{\epsilon})|\beta_i|)\right\rfloor}_{M_i},
\end{equation}
since $2^{m_i} {p_i^{\star}} $ is a rational integer: we have $M_i
- m_i + 1$ possible values for the integer $ 2^{m_i}
{p_i^{\star}}$. This means that we have $\prod_{i=0}^n (M_i - m_i
+ 1)$  polynomials candidates. If this amount is small enough, we
search for $p^\star$ by computing %exhaustively
the norms $ ||f -q ||_{\infty,[0,a]}$, $q$ running among the
possible polynomials. Otherwise, we need an additional step to
decrease the number of candidates.
%In some cases, the bounds given by \eqref{bounds} are in a too wide
%range.
% Hence, we give now a method that allows to refine the
%bounds.
Hence, we give now a method for this purpose.

 Condition \eqref{inflambda} means
\begin{equation} \label{ineqtr}
f(x) - \lambda \hat{\epsilon} \leq \sum_{i=0}^{n} {p_i^{\star}}
x^i \leq f(x) + \lambda \hat{\epsilon}
\end{equation}
for all $x \in [0,a]$. In particular, we have
$$
  f(0) - \lambda \hat{\epsilon} \leq p_0^{\star}
\leq f(0) + \lambda \hat{\epsilon}
$$
i.e., since $2^{m_0} p_0^{\star} $ is an integer,
$$
 \lceil 2^{m_0} ( f(0) - \lambda \hat{\epsilon}) \rceil \leq 2^{m_0} p_0^{\star}
\leq \lfloor  2^{m_0}  (f(0) + \lambda \hat{\epsilon}) \rfloor.
$$
The $n+1$ inequations given by \eqref{bounds} define a polytope to
which the numerators (i.e. the $2^{m_i} p_i^{\star}$)  belong. The
idea is to try to make this polytope smaller in order to reduce
our final exhaustive search. We do that thanks to inequations
\eqref{ineqtr} considered for a certain number (chosen by the
user) of values of $x \in [0,a]$. Once we got a small enough
polytope, we start our exhaustive search using libraries (such as
Polylib \cite{polylib} and CLooG \cite{cloog})
%the last one being
specially designed for scanning efficiently the integer points of
polytopes and producing only the corresponding loops in our
program of exhaustive search. CLooG implements the Quiller\'e et
al. algorithm \cite{QuiRaWil2000}.

\section{Examples}

We implemented in Maple a weakened version of the process
described in the previous section. By this, we mean that in the
step of refinement of the polytope, we only determine its vertices
using the simplex method instead of scanning its integer points.
The program first computes the bounds obtained from Chebyshev
polynomials and then, if these bounds are too large, computes the
vertices of the polytope obtained from  inequations
 \eqref{bounds} and inequations \eqref{ineqtr} considered for $x_i =
\frac{i}{d} A$ where $d$ is an integer parameter chosen by the
user, $i$ an integer, $ 0 \leq i \leq d$ and $A$ is a rational
number ``close'' and less than  or equal to $a$.

\subsection{Cosine function in $[0,\pi/4]$ with a
degree-$3$ polynomial}

In $[0,\pi/4]$, the distance between the cosine function and its
best degree-$3$ minimax approximation is $0.00011$. This means
that such an approximation is not good enough for single-precision
implementation of the cosine function. It can be of interest for
some special-purpose implementations. In this example, the bounds
given by the first step (associated to Chebyshev polynomials) are
good enough to avoid the use of the polytope refinement.

\begin{verbatim}
>m := [12,10,6,4]:polstar(cos,Pi/4,3,m);

  "minimax = ", .9998864206

         + (.00469021603 + (-.5303088665 + .06304636099 x) x) x

            "Distance between f and p =", .0001135879209

                            3   17  2     5
              "hatp = ", - x  - -- x  + ---- x + 1
                                32      1024


           "Distance between f and hatp =", .0006939707


>Do you want to continue (y;/n;)? y;
>Enter the value of parameter lambda: 1/2;

degree 0: 4 possible values between 2047/2048 and
          4097/4096
degree 1: 22 possible values between -3/512 and
          15/1024
degree 2: 5 possible values between -9/16 and
          -1/2
degree 3: 1 possible values between 1/16 and
          1/16
440 polynomials need be checked

>Do you want to try to refine the bounds (y;/n;)?n;


                            1  3   17  2    3      4095
             "pstar = ",   -- x  - -- x  + --- x + ----
                           16      32      512     4096


           "Distance between f and pstar =", .0002441406250

                "Time elapsed (in seconds) =", 1.840
\end{verbatim}

In this example, the distance between $f$ and $p^*$ is
approximately $0.35$ times the distance between $f$ and $\hat{p}$.
Using our method saves around $-\log_2(0.35) \approx 1.5$ bits of
accuracy.

\subsection{Exponential function in $[0,\log(1+1/2048)]$ with a
degree-$3$ polynomial}

In $[0,\log(1+1/2048)]$, the distance between the exponential
function and its best degree-$3$ minimax approximation is around
$1.8 \times 10^{-17}$, which should be sufficient for a faithfully
rounded double precision implementation provided there is much
care in the polynomial evaluation. The bounds given to get
$p^{\star}$ using the first step are too large (there are 18523896
polynomials to test). Hence, we must use the polytope refinement.

\begin{verbatim}
>Digits:=30:
>m := [56,45,33,23]: polstar(exp,log(1.+1./2048),3,m);

"minimax = ", .999999999999999981509827946165 +
(1.00000000000121203815619648271
 + (.499999987586063030320493910112
 + .166707352549861488779274879363 x) x) x

                                                   -16
    "Distance between f and p =", .1849017208895 10


             1398443  3   4294967189  2   35184372088875
  "hatp = ", ------- x  + ---------- x  + -------------- x
             8388608      8589934592      35184372088832


           72057594037927935
         + -----------------
           72057594037927936


    "Distance between f and hatp =",
                                            -16
                 .23624220969326235229443 10

>Do you want to continue (y;/n;)? y;
>Enter the value of parameter lambda: 1;

degree 0: 6 possible values between
18014398509481983/18014398509481984
          and 72057594037927937/72057594037927936
degree 1: 109 possible values between
35184372088821/35184372088832
          and 35184372088929/35184372088832
degree 2: 146 possible values between 4294967117/8589934592
          and 2147483631/4294967296
degree 3: 194 possible values between 699173/4194304
          and 1398539/8388608
18523896 polynomials need be checked

>Do you want to try to refine the bounds (y;/n;)?y;
>Enter the value of parameter d: 25;

degree 0: 2 possible values between
72057594037927935/72057594037927936
          and 1
degree 1: 27 possible values between 35184372088857/35184372088832
          and 35184372088883/35184372088832
degree 2: 32 possible values between 536870897/1073741824
          and 4294967207/8589934592
degree 3: 44 possible values between 1398421/8388608
          and 21851/131072
76032 polynomials need be checked


>Do you want to try to refine the bounds (y;/n;)?n;
>Do you want to change the value of Digits (y;/n;)?y;
>Enter the value of Digits: 21;

            1398443  3   2147483595  2   35184372088873
"pstar = ", ------- x  + ---------- x  + -------------- x
            8388608      4294967296      35184372088832

           72057594037927935
         + -----------------
           72057594037927936

    "Distance between f and pstar =",
                                            -16
                 .20246280367096470182285 10

         "Time elapsed (in seconds) =", 54721.961
\end{verbatim}

In this last example, the distance between $f$ and $p^*$ is
approximately $0.85$ times the distance between $f$ and $\hat{p}$.
Using our method saves around $-\log_2(0.85) \approx 0.22$ bits of
accuracy.

\bibliographystyle{plain}

\end{document}